\documentclass[preprint,12pt]{elsarticle}

\usepackage{amssymb}
\usepackage{amsmath}
\usepackage{caption}
\usepackage{booktabs}
\usepackage{multirow}

\journal{Physica A: Statistical Mechanics and its Applications}

\begin{document}

\begin{frontmatter}

\title{Various Vicsek Models with Underlying Network Characteristics}

\author[label1]{Haoshuai Wang}
\author[label1]{Zhaoqi Dong}
\author[label2]{Lei Chen\corref{cor1}}
\cortext[cor1]{Corresponding author. Email: bit$\_$chen@bit.edu.cn}
\affiliation[label1]{organization={School of Automation, Beijing Institute of Technology},
            addressline={5 South Zhongguancun Street}, 
            city={Beijing},
            postcode={100081}, 
            country={China}}

\affiliation[label2]{organization={School of Artificial Intelligence, Beijing Institute of Technology},
            addressline={5 South Zhongguancun Street}, 
            city={Beijing},
            postcode={100081}, 
            country={China}}

\begin{abstract}
Collective motion is a fundamental phenomenon in biological swarms. As a framework for studying synchronization in motions, the Vicsek model is simple and efficient, assuming isotropic interactions with a complete field of view. Drawing inspiration from natural swarms, we incorporate realistic constraints into the model. By analysing the interaction structures from the complex network perspective, we demonstrate that models with the homogeneous interaction rules naturally form Erd\H{o}s-R\'enyi networks, whereas the introduction of heterogeneity leads to Barab\'asi-Albert networks. Furthermore, we discover that the model's synchronization is fundamentally governed by the average degree of the interaction network. Through a comparative analysis across these topologies, we identify a stretched-exponential relationship between the average degree and the synchronization metrics.
\end{abstract}


\end{frontmatter}

\section{Introduction}
\label{sec1}

In nature, collective motions are frequently observed, including the coordinated swimming of fish, the flocking of birds in flight, and the large-scale movements of insect swarms~\cite{vicsek2012collective, krongauz2022vision, bailo2022review}. Among these motion patterns, spontaneous synchronization, where individuals adjust their direction of motion based on local information and the swarm eventually moves in a common direction, is extensively studied in physics, biology and engineering~\cite{clusella2021phase, olfati2007consensus}. Subsequent studies seek to uncover the mechanisms that enable local interactions to produce large-scale synchronization~\cite{jin2025role, you2023modified}. Beyond its role in explaining natural swarm systems, the study of mechanisms also provides essential insights into the design of engineered multi-agent systems~\cite{ren2005second, jadbabaie2003coordination, kia2019tutorial}.

A widely used framework for investigating such mechanisms is the Vicsek model proposed in 1995~\cite{vicsek1995novel}. Due to its simple rules and rich collective behaviors, the Vicsek model becomes a standard reference for understanding synchronization~\cite{gregoire2004onset, zhang2021complexnoise}. It assumes that individuals interact through isotropic rules and play identical roles within the swarm~\cite{czirok1997spontaneously, yang2018adaptive}. However, in real biological systems, the view of animals is not omnidirectional, and swarms often exhibit hierarchical heterogeneity. It is thus more reasonable to assume limited view angles and heterogeneity in swarm models to better mimic the real collective behaviors~\cite{chen2022direction,lu2023remote}. Tian et al.~\cite{tian2009optimal} extend the Vicsek model to incorporate limited sensing sectors. Li et al.~\cite{li2011optimal} investigate how anisotropic perception influences collective dynamics, revealing that the angular range of interaction plays a critical role in synchronization. Notably, studies on the optimal view angle of restricted-view variants~\cite{shirazi2018optimal,zhang2023optimal} report that a view angle of approximately $3\pi/2$ yields the fastest and strongest synchronization. These findings suggest that introducing a blind spot helps agents discard redundant or inconsistent information from the rear, thereby making the local signals used for alignment more focused and effective. Parallel to these geometric constraints, a notable discovery is that Nagy et al.~\cite{nagy2010hierarchical} reveal a well-defined hierarchy of directional influence, in which changes in heading tend to propagate from a small subset of pigeons to the rest of the flock. How these restricted fields of view and heterogeneity jointly shape the interaction structure and synchronization remains to be further explored.

To deeply understand how these interaction rules translate into synchronization, it is essential to examine the underlying structure of information exchange~\cite{yang2023connectivity, kumar2021efficient, rakshit2024stability}. In the Vicsek model, the continuous update of headings based on the states of neighbors can be modeled as a dynamic complex network, where particles represent nodes and interaction links serve as edges~\cite{lu2022neighbors, sun2022structure}. From this perspective, synchronization of the swarm is essentially a process of information diffusion over the network, so the topological features of networks play a crucial role in determining the global alignment. However, for Vicsek-type systems with restricted fields of view and hierarchical heterogeneity, it remains unclear how these biological constraints reshape the statistical properties of the interaction networks. This motivates us to explicitly characterize the network structures generated by different sensing mechanisms and to explore the relationship between average degree and synchronization metrics.

The paper is organized as follows. In Sec.~\ref{sec2}, we introduce Vicsek-type variants with heterogeneous sensing and restricted fields of view, and formulate their interaction networks and degree-based characterization. Sec.~\ref{sec3} describes the simulation setup and presents the numerical results, including a systematic analysis of how the final synchronization level depends on the initial average degree and its stretched-exponential fit. Finally, Sec.~\ref{sec4} concludes the paper and discusses implications for both biological swarms and engineered multi-agent systems.

\section{Models and discovery}
\label{sec2}

In this section, we outline the Standard Vicsek Model (SVM) and extend it to include restricted fields of view and heterogeneity. Furthermore, we interpret the particle interactions as dynamic networks, exploring how different local interaction rules shape the underlying network topology.

\subsection{The Vicsek Model and Variants}
In the SVM, a group of $N$ particles move at a constant speed $v_0$ in a $L \times L$ square. The state of particle $i$ can be described by the position $\boldsymbol{x}_i$ and the direction $\theta_i$ which is randomly distributed within $[0, 2\pi]$ at the initial time. The particle $i$ moves according to the laws as follows:
\begin{eqnarray}
  \boldsymbol{x}_i(t+1)  & = & \boldsymbol{x}_i(t) + {{\boldsymbol{v}}_i(t)}\;T \nonumber \\
  \theta_i(t+1) & = & {arg}({\boldsymbol{U}_i}) + {\xi}_i(t),
  \label{eq:pos_update}
\end{eqnarray}
where $T$ is the time step and ${\xi}_i(t)$ is the noise uniformly distributed in $[-\eta/2, \eta/2]$. The velocity vector is defined as ${{\boldsymbol{v}}_i(t)} = ({v_0}\cos{\theta_i}, {v_0}\sin{\theta_i})$. The vector ${\boldsymbol{U}_i}$ represents the sum of the motion vectors of particle $i$'s neighbors, which is defined by
\begin{eqnarray}
  {\boldsymbol{U}_i} = {\sum\limits_{j \in {{\cal N}_i^{\mathrm{SVM}}}(t)} {{{\boldsymbol{v}}_j}} (t)},
\end{eqnarray}
where $\mathcal{N}_i^{\mathrm{SVM}}(t) = \{ j \mid \|\boldsymbol{x}_j - \boldsymbol{x}_i\| \leqslant r \}$ denotes the set of neighbors.

\begin{figure}[htbp]
    \centering
    
    \begin{minipage}[t]{0.45\textwidth}
        \centering
        \includegraphics[width=\linewidth]{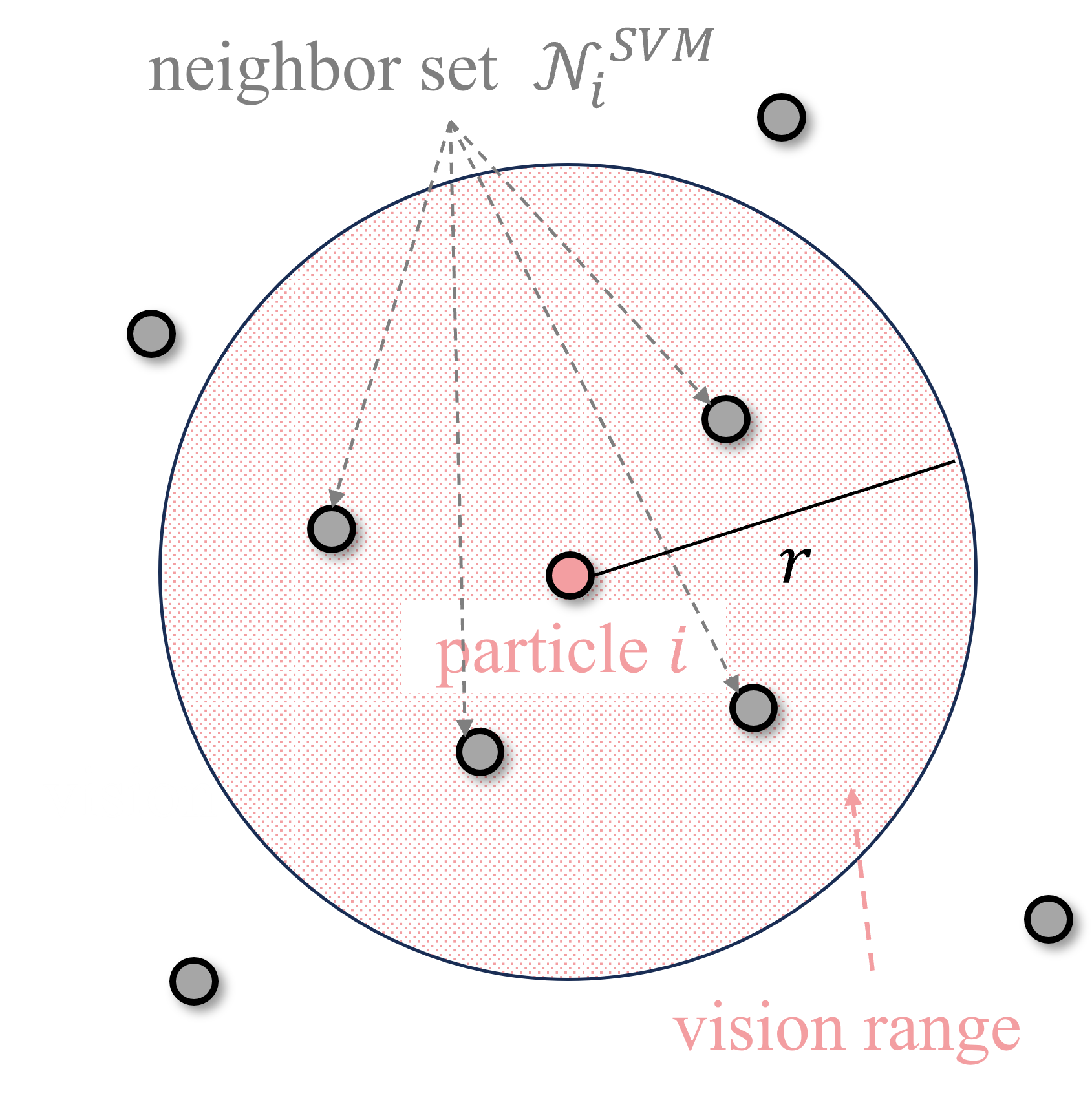}
        \caption*{(a) SVM}
    \end{minipage}
    \hfill
    \begin{minipage}[t]{0.45\textwidth}
        \centering
        \includegraphics[width=\linewidth]{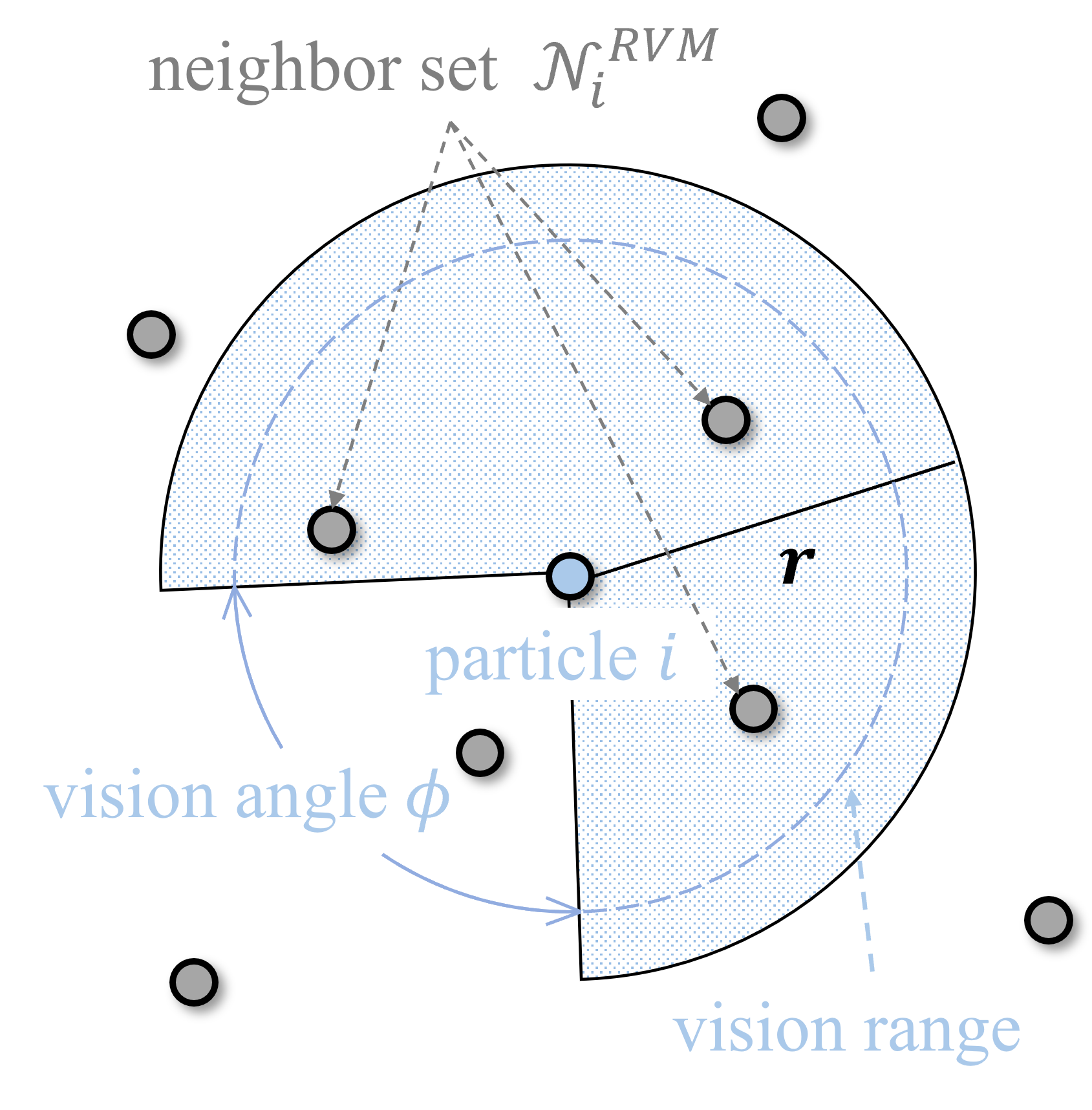}
        \caption*{(b) RVM}
    \end{minipage}

    \caption{Schematic illustration of the local interaction rules. (a) SVM: The focal particle $i$ interacts isotropically with all neighbors within a circular radius $r$. (b) RVM: The interaction is restricted to a visual sector defined by the radius $r$ and the vision angle $\phi$. The shaded regions represent the effective perception field of particle $i$.}
    \label{fig:ER_models}
\end{figure}

The field of perception for every particle in SVM is a complete disk 2D case characterized only by its sensor radius $r$. The Restricted-view-angle Vicsek model (RVM) are introduced to investigative the influences of view angles in the swarm synchronization. The only difference from the SVM is that the mechanism of neighbors selection which can be described by
\begin{eqnarray}
  {\mathcal{N}_i^{\mathrm{RVM}}(t)} = \left\{ j \,\middle|\, \big\lVert \boldsymbol{x}_j(t) - \boldsymbol{x}_i(t) \big\rVert \leqslant r,\ \lvert \vartheta_{ij}(t) \rvert \leqslant \phi/2 \right\},
\end{eqnarray} 
where $\phi \in (0,2\pi]$ denotes the view angle, and $\vartheta_{ij}(t)$ denotes the angle between the heading direction of particle $i$ and the relative position vector $\boldsymbol{x}_j(t) - \boldsymbol{x}_i(t)$. When $\phi = 2\pi$, the RVM becomes SVM.

Not only the limit view angles, but also the heterogeneous one exist in the animals. Guided by these phenomena, we introduce two model variants. In each variant, a fraction $\gamma \in (0,1)$ of the population has enhanced interaction capabilities characterized by an extended interaction radius $\sigma r$ ($\sigma > 1$). We denote this subset as the heterogeneous particles $\mathcal{H}$ with cardinality $|\mathcal{H}| = \gamma N$, while the remaining particles are referred to as ordinary particles.

\paragraph{Leader Perception Vicsek Model (LPVM)}
In this variant, the heterogeneous particles possess a superior sensory range. The interaction radius depends on the identity of the observer (particle $i$). We define the perception radius $R_i$ as
\begin{eqnarray}
    R_i = 
    \begin{cases} 
        \sigma r, & \text{if } i \in \mathcal{H}, \\ 
        r, & \text{if } i \notin \mathcal{H}. 
    \end{cases}
\end{eqnarray}
The set of neighbors is thus defined as
\begin{eqnarray}
\mathcal{N}_i^{\mathrm{LPVM}}(t) = \left\{ j \,\middle|\, \big\lVert \boldsymbol{x}_j(t) - \boldsymbol{x}_i(t) \big\rVert \leqslant R_i,\ \lvert \vartheta_{ij}(t) \rvert \leqslant \phi/2 \right\}.
\label{eq:lpvm_neighborhood}
\end{eqnarray}
Here, heterogeneous particles can integrate information from more distant neighbors, while ordinary nodes remain restricted to the standard range.

\begin{figure}[htbp]
    \centering

    \begin{minipage}[t]{0.45\textwidth}
        \centering
        \includegraphics[width=\linewidth]{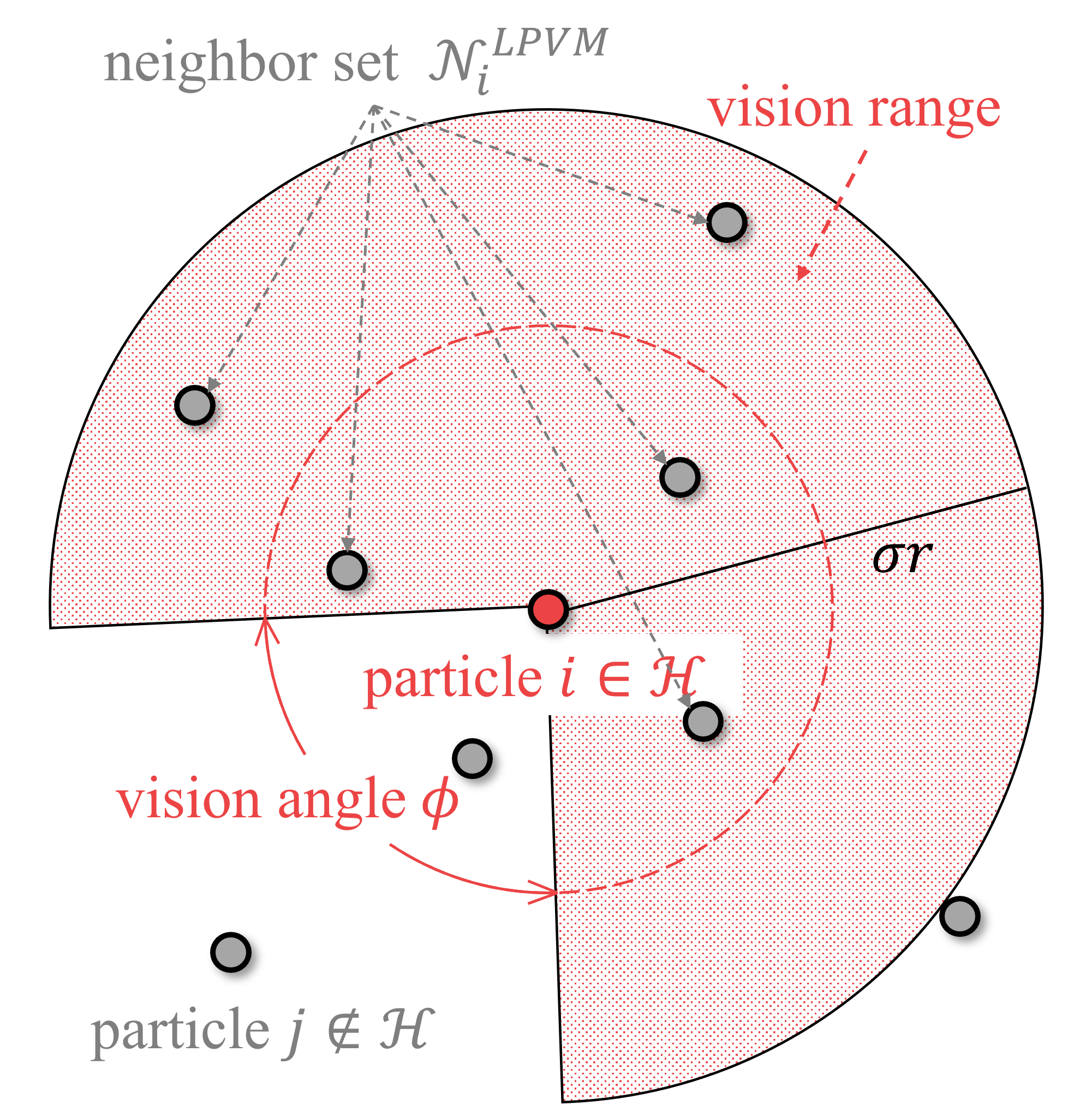}
        \caption*{(a) LPVM}
    \end{minipage}
    \hfill
    \begin{minipage}[t]{0.45\textwidth}
        \centering
        \includegraphics[width=\linewidth]{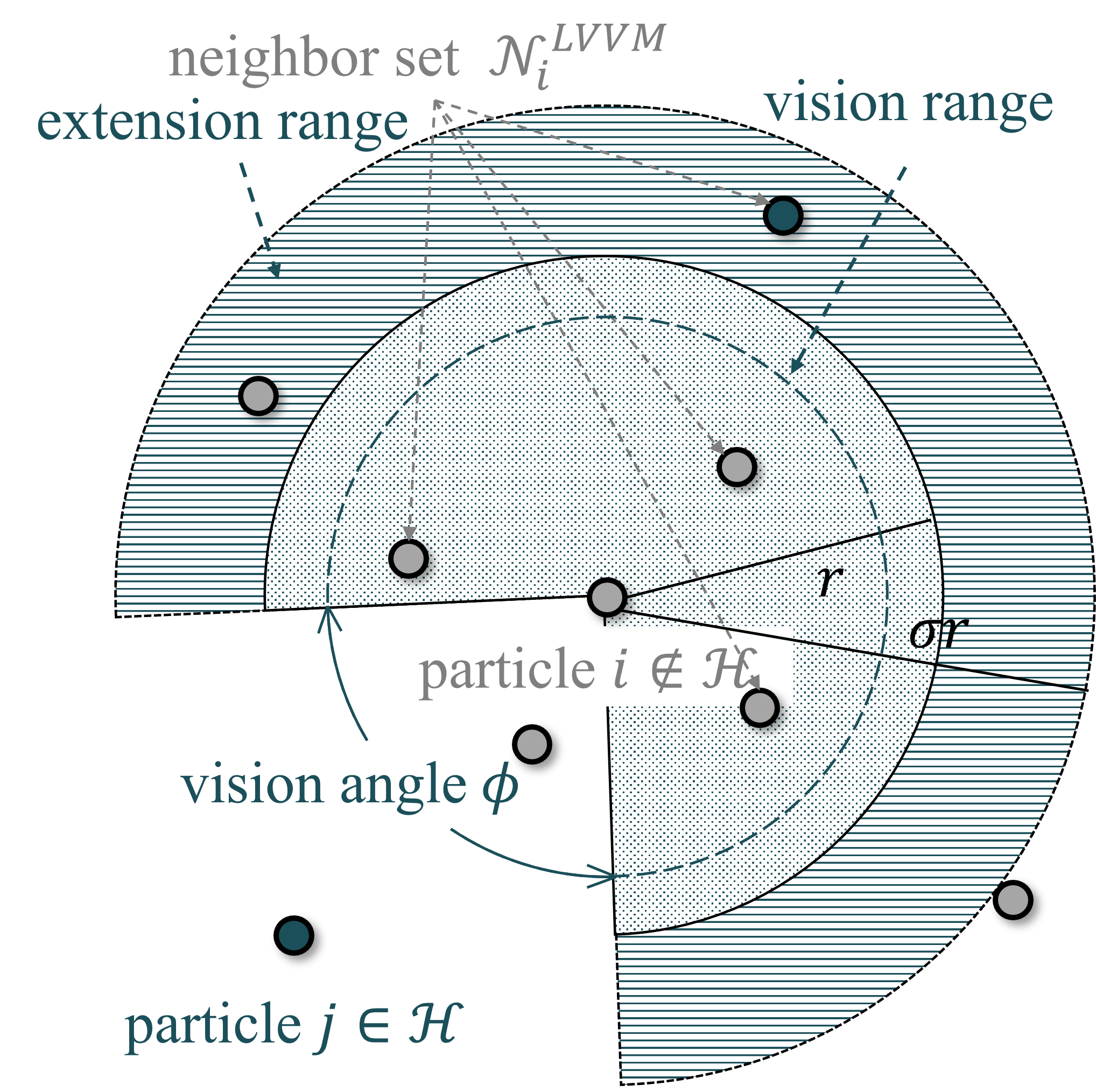}
        \caption*{(b) LVVM}
    \end{minipage}

    \caption{Schematic illustration of the heterogeneous interaction mechanisms. (a) LPVM: A heterogeneous particle $i \in \mathcal{H}$ (red node) possesses an extended perception range $\sigma r$, enabling it to sense distant neighbors. (b) LVVM: Heterogeneous neighbors $j \in \mathcal{H}$ (dark blue nodes) possess an extended visibility range, allowing them to be detected by the focal particle within the extended radius $\sigma r$.}
    \label{fig:BA_models}
\end{figure}

\paragraph{Leader Visibility Vicsek Model (LVVM)}
Conversely, in the LVVM, the heterogeneous particles are more influential. The interaction radius depends on the identity of the neighbor (particle $j$). The effective visibility radius $R_j$ is defined as
\begin{eqnarray}
    R_j = 
    \begin{cases} 
        \sigma r, & \text{if } j \in \mathcal{H}, \\ 
        r, & \text{if } j \notin \mathcal{H}. 
    \end{cases}
\end{eqnarray}
The neighborhood implies that a neighbor $j$ is considered only if it falls within the observer's field of view and its specific visibility range
\begin{eqnarray}
\mathcal{N}_i^{\mathrm{LVVM}}(t) = \left\{ j \,\middle|\, \big\lVert \boldsymbol{x}_j(t) - \boldsymbol{x}_i(t) \big\rVert \leqslant R_j,\ \lvert \vartheta_{ij}(t) \rvert \leqslant \phi/2 \right\}.
\label{eq:lvvm_neighborhood}
\end{eqnarray}
In this scenario, all particles share the same sensing capability, but heterogeneous nodes can influence others from a greater distance, effectively acting as long-range beacons. 

\subsection{Discovery from complex network perspective}

In the Vicsek framework, local interactions correspond to particles continuously collecting information from their neighbours and updating their own motion, which can be viewed as an information exchange process among particles. To capture how this information propagates through the swarm, we adopt a complex network description, in which particles are represented as nodes and interaction links as edges.

\begin{figure}[t]
    \centering

    \begin{minipage}[t]{0.49\textwidth}
        \centering
        \includegraphics[width=\linewidth]{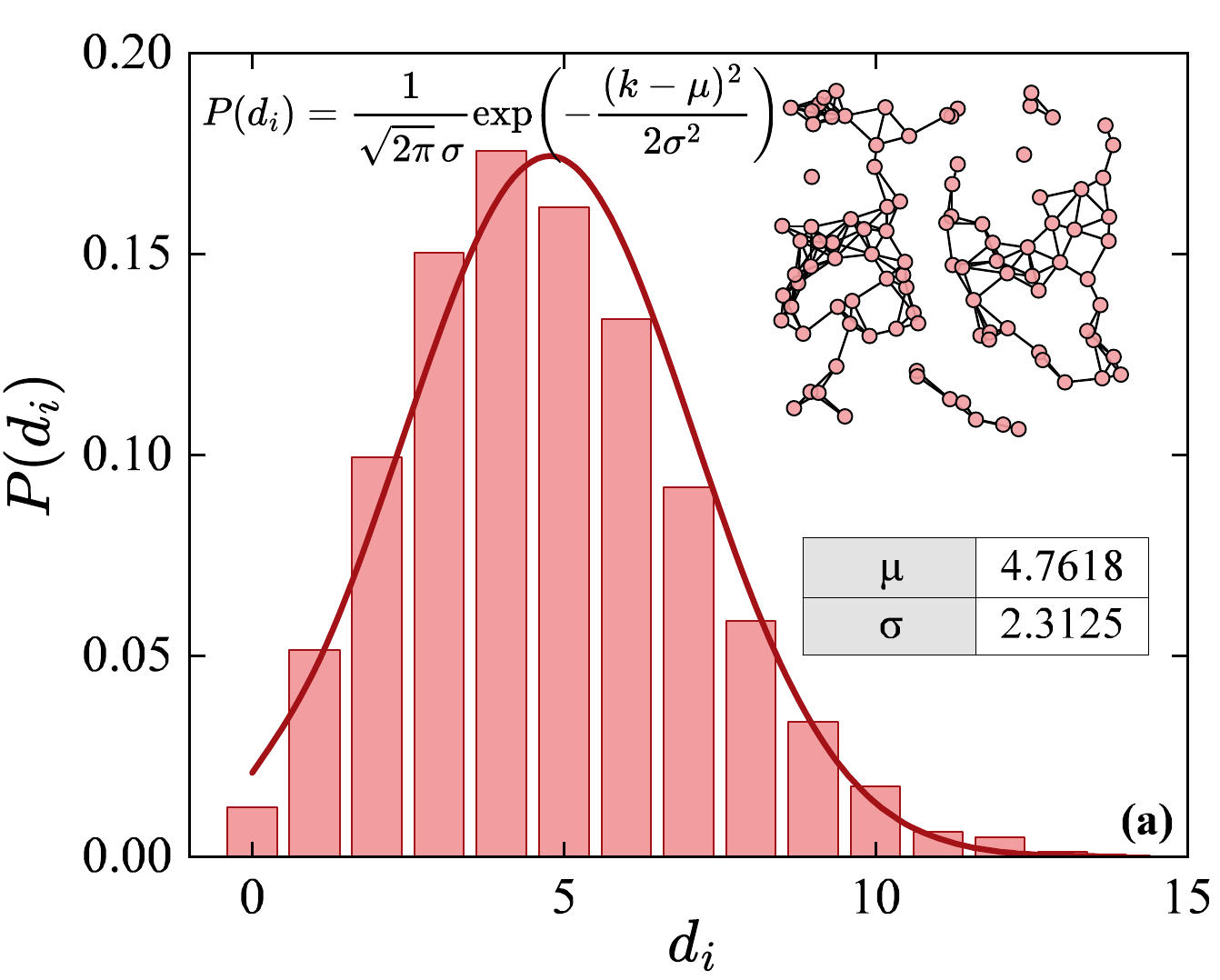}
    \end{minipage}
    \hfill
    \begin{minipage}[t]{0.49\textwidth}
        \centering
        \includegraphics[width=\linewidth]{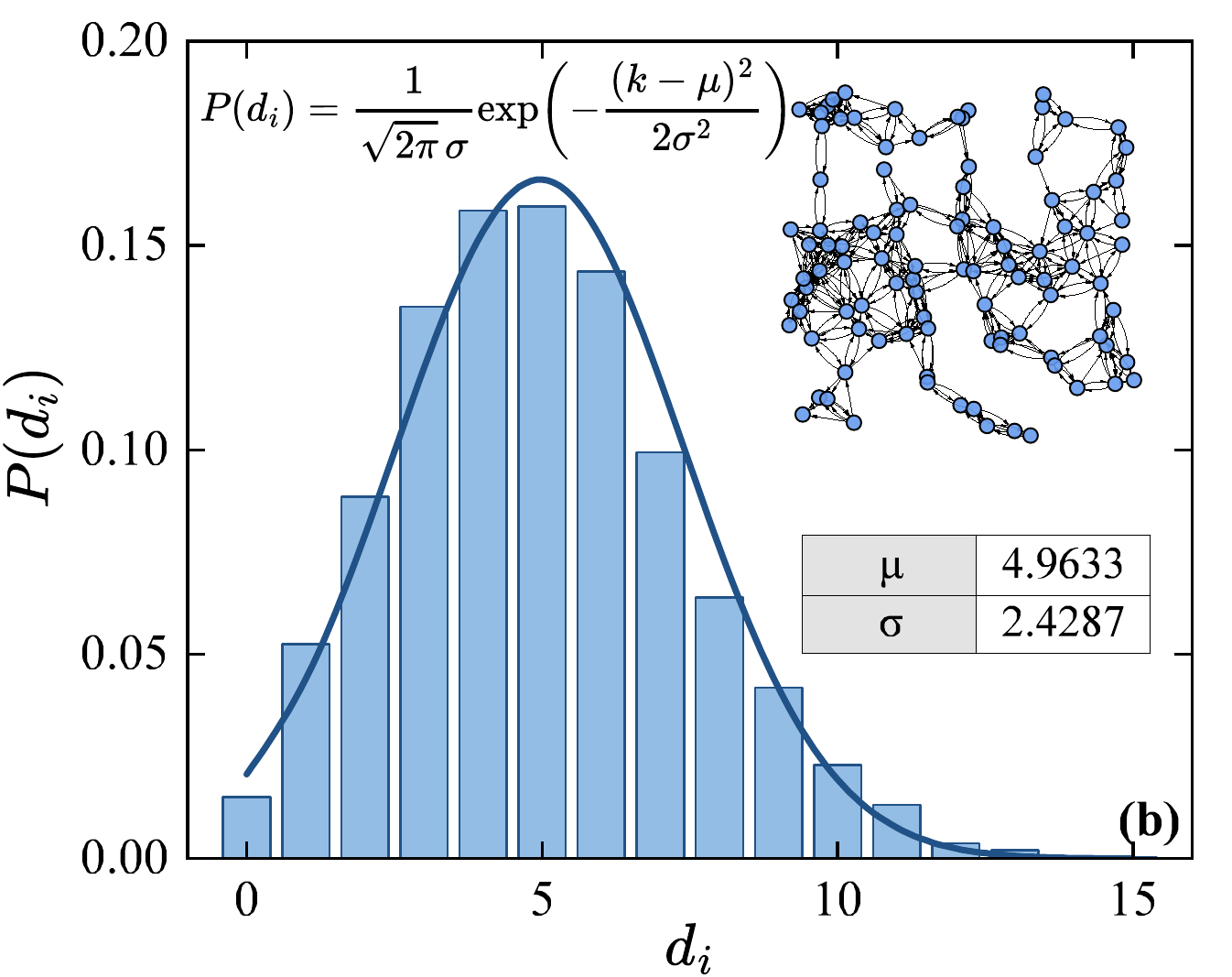}
    \end{minipage}

    \vspace{0.1cm} 
    
    \begin{minipage}[t]{0.49\textwidth}
        \centering
        \includegraphics[width=\linewidth]{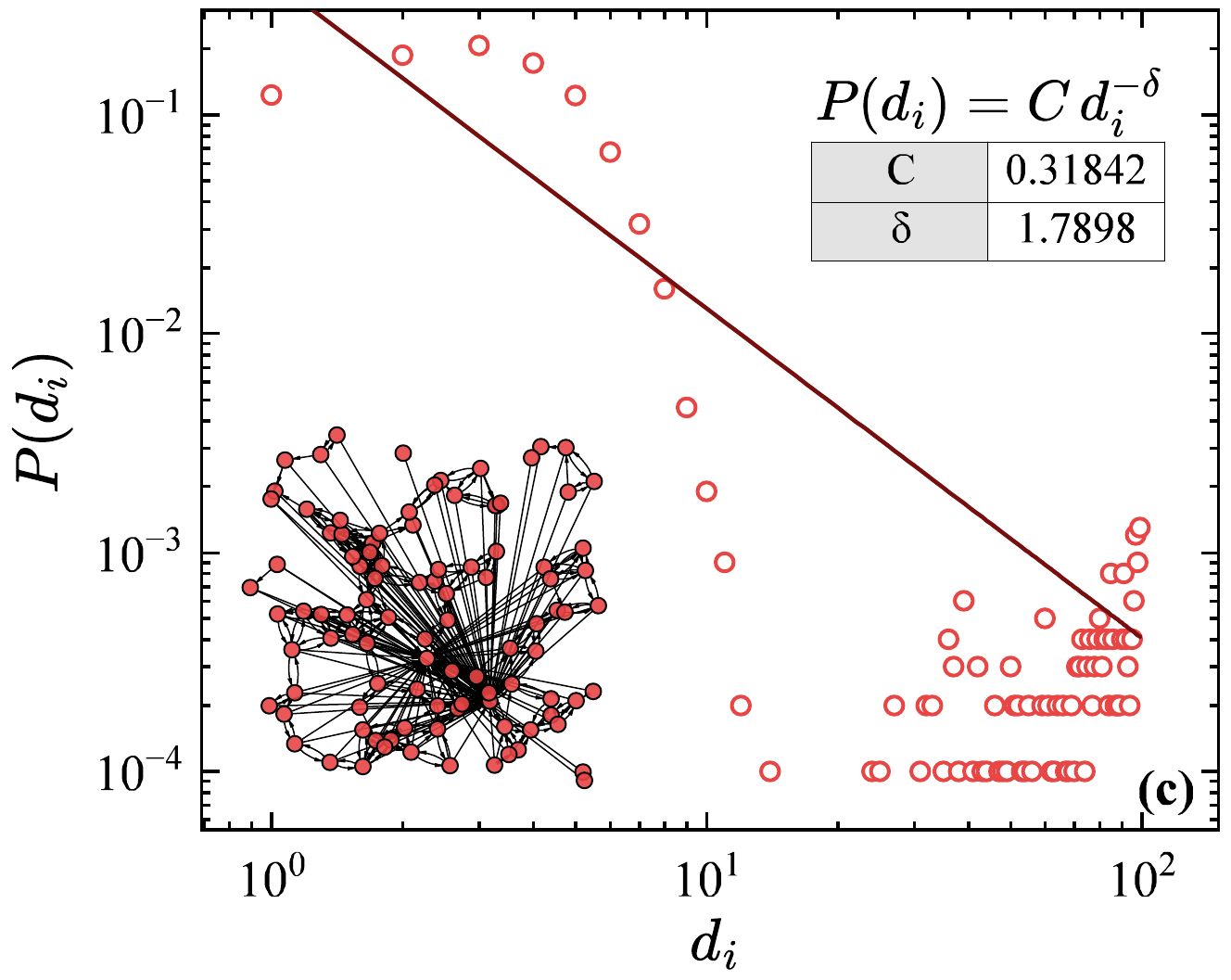}
    \end{minipage}
    \hfill
    \begin{minipage}[t]{0.49\textwidth}
        \centering
        \includegraphics[width=\linewidth]{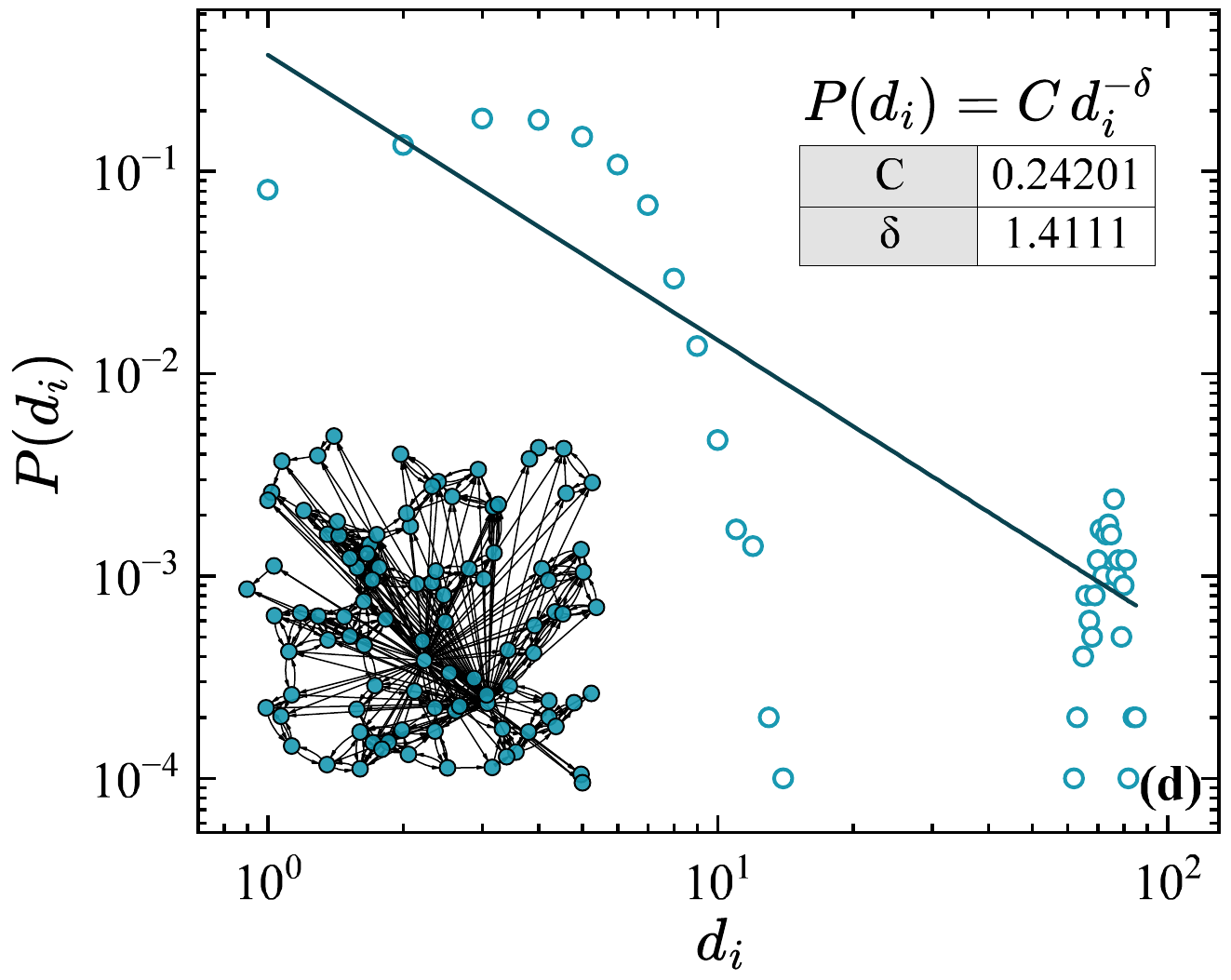}
    \end{minipage}

    \caption{Degree distributions of the interaction networks. \textbf{(a)} SVM and \textbf{(b)} RVM display approximately normal degree distributions, fitted by Gaussian curves. In contrast, \textbf{(c)} LPVM and \textbf{(d)} LVVM exhibit heavy-tailed power-law distributions $P(d_i) \sim d_i^{-\delta}$ on log-log scales, characteristic of heterogeneous scale-free networks. The insets illustrate typical network snapshots. Parameters: $\langle d \rangle = 5.0$, $N = 100$, domain size $10 \times 10$.}
    \label{fig:degree_distributions}
\end{figure}

From the complex network perspective, the interaction structure at a given time can be represented as a graph $\mathcal{G} = (\mathcal{N}, \mathcal{E})$, where $\mathcal{N} = \{ n_k \mid k = 1,2,\dots,N \}$ is the set of nodes and $N$ is the number of particles in the swarm, and $\mathcal{E} \subseteq \mathcal{N} \times \mathcal{N}$ is the set of edges that represent neighbour relationships between nodes. We use $a_{ij} = 1$ to indicate that particle $i$ is a neighbour of particle $j$ and that information flows from $i$ to $j$, while $a_{ij} = 0$ denotes the absence of such a connection. The adjacency matrix $\mathbf{A}$ of a graph $\mathcal{G}$ with $N$ nodes serves as a fundamental representation of the network topology. Among the many characteristics of graph, node degree provides the most direct measure of how many neighbours a particle can exchange information with and thus of its local influence in the interaction network. For the SVM, the interaction network is undirected and the adjacency matrix is symmetric. In this case the degree of node $i$ is
\begin{eqnarray}
d_i = \sum_{j=1}^{N} a_{ij}.
\label{eq:degree_undirected}
\end{eqnarray}
For the variants, the interaction network is directed due to asymmetric perception ranges. Since each particle's motion is determined by the information it receives from its neighbors, in-degree correlates with swarm dynamics. In subsequent studies, all analyses concerning degree in directed networks refer to in-degree.

Since the information propagation is constrained by who is connected to whom, a natural starting point is to examine the topology of the interaction network, and in particular the distribution of node degrees. Fig.~\ref{fig:degree_distributions} shows that the interaction networks generated by the SVM and RVM are close to homogeneous random graphs with narrowly peaked degree distributions, whereas the LPVM and LVVM produce heterogeneous networks with heavy-tailed power-law degree statistics characteristic of scale-free structures. Under the information-transfer interpretation, Erd\H{o}s-R\'enyi(ER)-like networks support a relatively uniform diffusion of alignment information, while in Barab\'asi-Albert(BA)-like networks with heavy-tailed degree distributions a few highly connected nodes are expected to dominate the communication process and facilitate faster global alignment. In the next subsection we quantify this intuition by analysing how the average degree of the interaction network correlates with the macroscopic order parameter for the four models. 

\section{Results and discussions}
\label{sec3}

In order to analyse the laws between the average degree and synchronization, we consider a system of \(N\) self-propelled particles moving in a two-dimensional plane. To better approximate real world swarms, we adopt open a square domain of size \(10 \times 10\) with open boundary conditions. The particles move with randomly generated initial positions and headings. The constant speed is fixed to \(v_0 = 0.03\). We explore four system sizes, \(N = 50, 100, 250, 500\), and iterate the dynamics up to \(T_{\max} = 500\) time steps for each realization. The fraction of heterogeneous particles is varied as \(\gamma = 0.1, 0.3, 0.5\), and the sensing range scaling factor that controls their enhanced perception is chosen from \(\sigma = 5, 10, 15, 20\). For the models with angular restriction, the visual angle is set to \(\phi = 3\pi/2\)~\cite{shirazi2018optimal}. The initial average degree of the interaction network, \({\left\langle d \right\rangle _{i}}\), is used as the main structural control parameter and is scanned over the range \(1, 2, \dots, 15\).

The average degree is used as a structural control parameter to quantify the connectivity of the interaction network. At the initial time, it is calculated as
\begin{equation}
\langle d\rangle_i=\frac{1}{N}\sum_{i=1}^Nd_i.
\end{equation}
To prescribe a target initial connectivity, the perception radius $r$ is iteratively adjusted until the measured average degree $\langle d \rangle_i$ matches the target $d_t$ within a tolerance of $0.1$. To discuss the synchronization of particles quantitatively, the order parameter is defined as
\begin{equation}
v_a=\frac{1}{Nv_0}\left|\sum_{i=1}^N\mathbf{v}_i\right|.
\end{equation}
This quantity is close to zero when particle headings are randomly distributed and approaches unity when the swarm moves coherently in a common direction. In the following, we focus on the stationary value of \(v_a\) obtained after sufficiently long timesteps.

\begin{figure}[ht]
    \centering
    \includegraphics[width=0.98\textwidth]{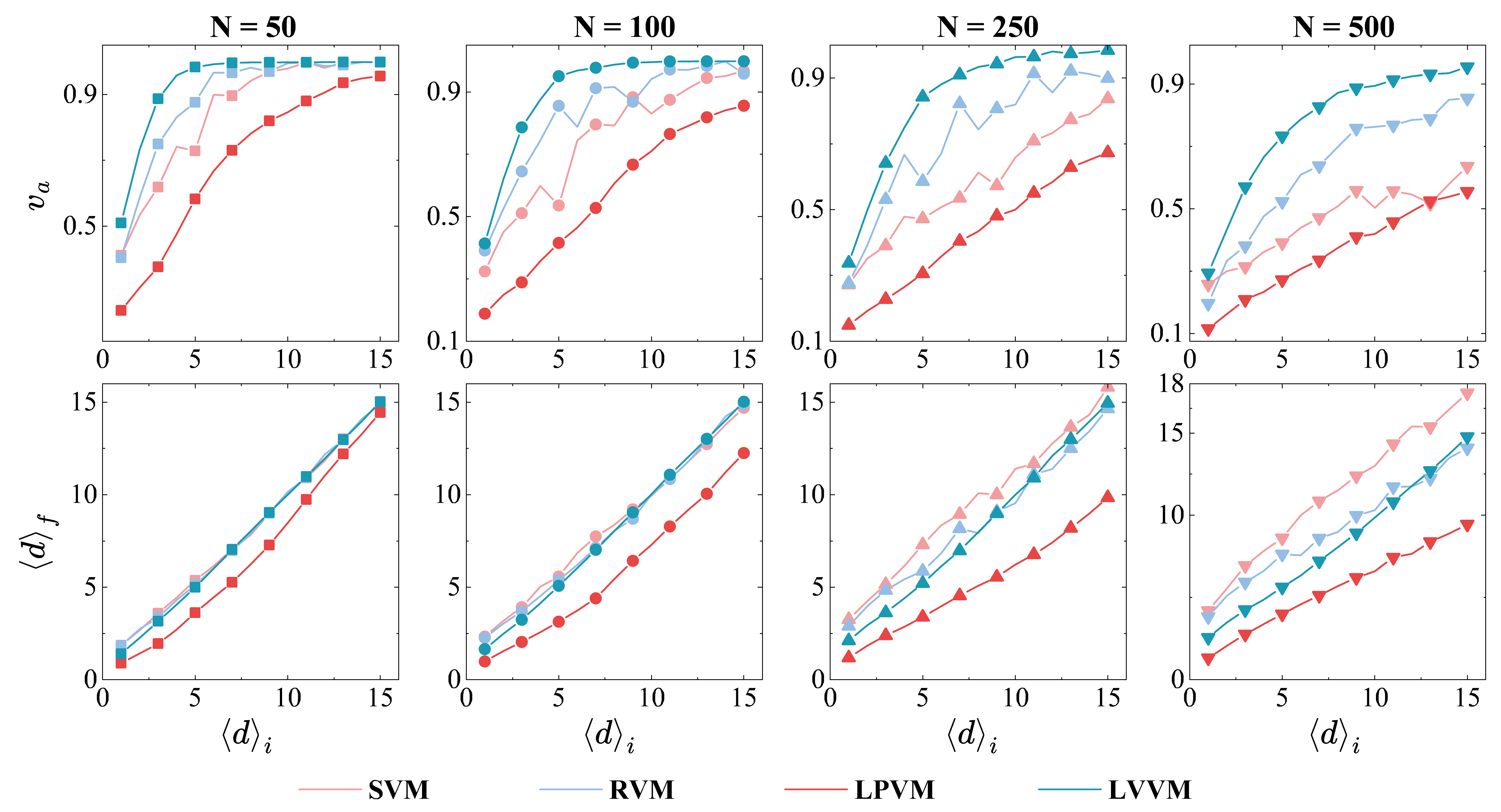}%
    \caption{Impact of initial connectivity on synchronization and network evolution. \textbf{Top row:} The stationary synchronization order parameter $v_a$ as a function of the initial average degree $\langle d \rangle_i$. \textbf{Bottom row:} The final average degree $\langle d \rangle_f$ versus the initial value $\langle d \rangle_i$. Columns correspond to different swarm sizes $N = 50, 100, 250, 500$.}
    \label{fig:di_vadf}
\end{figure}

\begin{figure}[t]
    \centering
    \includegraphics[width=0.98\textwidth]{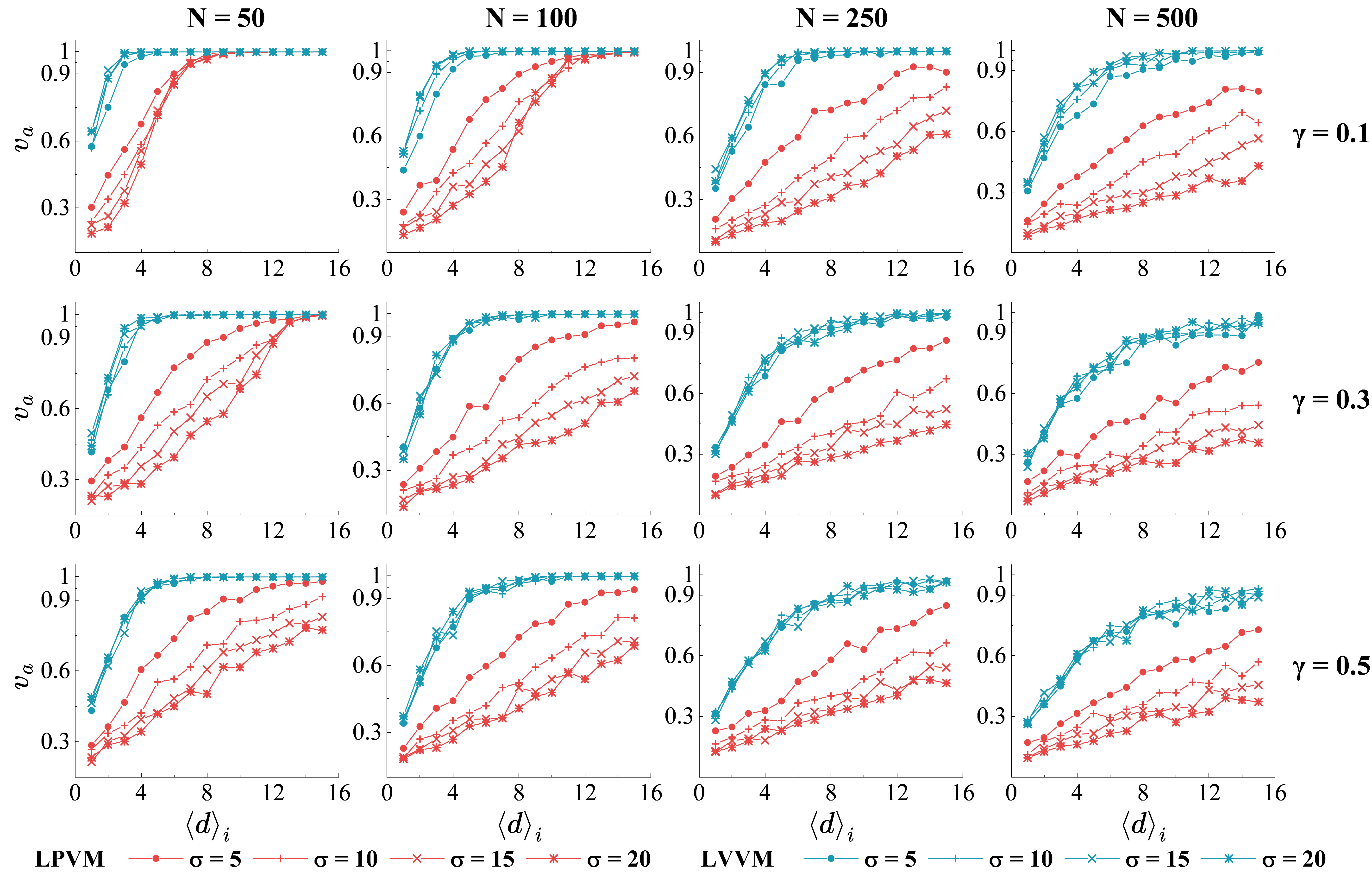}%
    \caption{Final synchronization level $v_a$ as a function of the initial average degree ${\left\langle d \right\rangle _{i}}$ for LPVM (red curves) and LVVM (blue curves). Columns correspond to different system sizes $N$, rows to different fractions of heterogeneous agents $\gamma$, and line styles to different sensing factors $\sigma$.}
    \label{fig:gamma_diva}
\end{figure}

To disentangle the roles of initial connectivity and system size in shaping the final interaction pattern, we first examine how the prescribed initial average degree ${\left\langle d \right\rangle _{i}}$ is translated into the final average degree ${\left\langle d \right\rangle _{f}}$ in the stationary regime. As can be seen from Fig.~\ref{fig:di_vadf}, as the number of particles increases, the average degree required to achieve synchronization also increases. Moreover, the average degree of the initial and final networks remained essentially unchanged. Among these models, the LPVM and LVVM demonstrates completely different synchronization effects. We next analyze the influence of the fraction of heterogeneous agents $\gamma$ and the sensing factor $\sigma$, which jointly tune the strength and prevalence of heterogeneity in the interaction network in Fig.~\ref{fig:gamma_diva}. For the LPVM, increasing either $\gamma$ or $\sigma$ systematically shifts the curves downwards and reduces their slope. A larger proportion of agents with increased in-degree and a stronger asymmetry in the sensing range therefore deteriorate global synchronization. In contrast, the LVVM remains close to full synchronization over a broad range of parameters. The dependence on $\gamma$ is weak and the curves for different $\sigma$ almost collapse, especially for large $N$. These results demonstrate that the visibility-based interaction rule is robust to the introduction of heterogeneous agents, and that further enlarging their influencing radius yields only limited additional gains in effective connectivity and synchronization.

\begin{figure}[ht]
    \centering
    \includegraphics[width=0.98\textwidth]{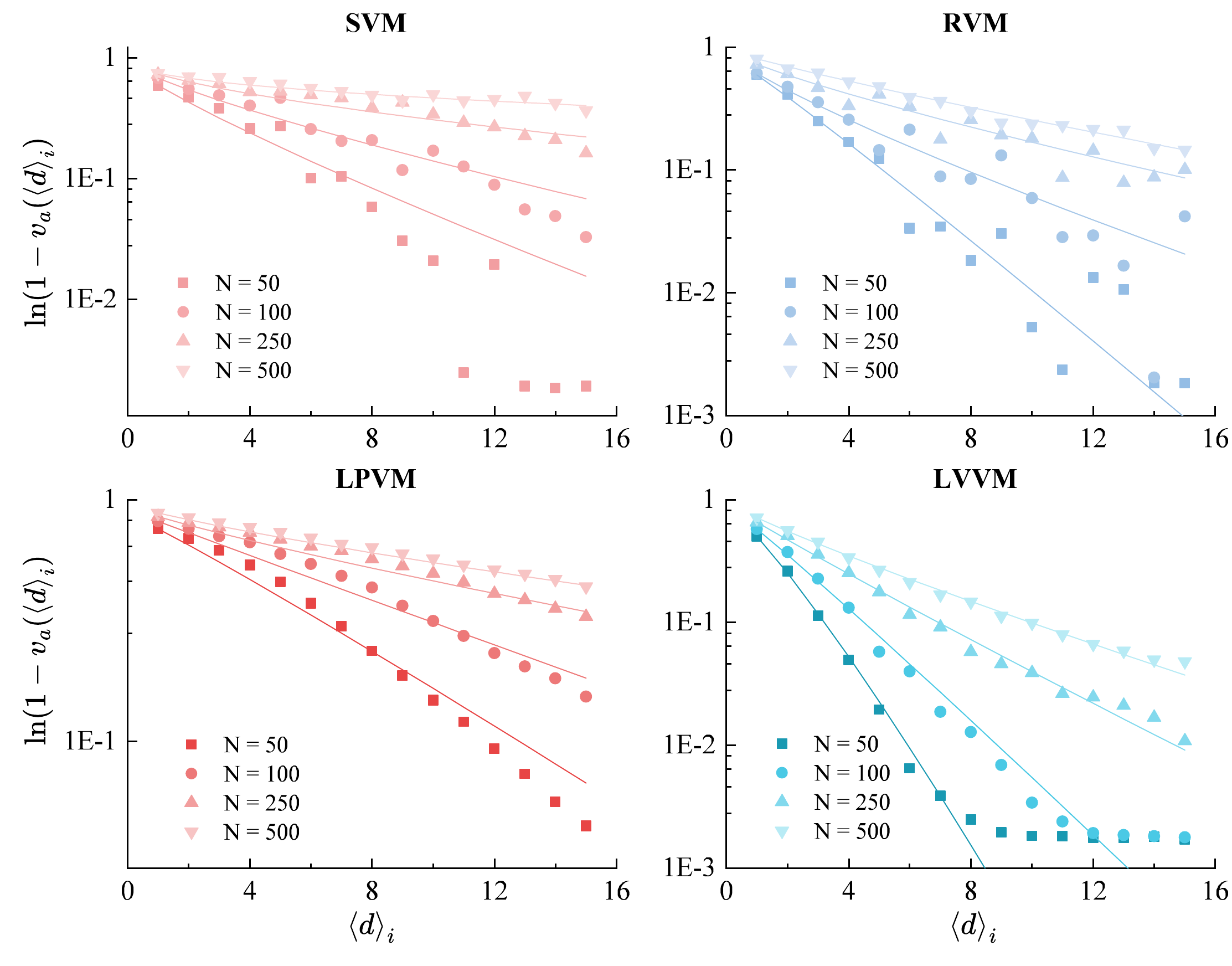}%
    \caption{Semi-log plots of $1-v_a$ versus the initial average degree ${\left\langle d \right\rangle _{i}}$ for the four models, where symbols denote simulation results and solid lines the corresponding stretched-exponential fits.}
    \label{fig:strexp_fits}
\end{figure}

\begin{table}[t]
\centering
\small
\begin{tabular}{llcccc}
\toprule
Model & Param & $N=50$ & $N=100$ & $N=250$ & $N=500$ \\
\midrule
\multirow{4}{*}{SVM}
  & $a$    & 0.875 & 0.911 & 0.943 & 0.961 \\
  & $b$    & 0.396 & 0.297 & 0.258 & 0.258 \\
  & $c$    & 0.857 & 0.800 & 0.638 & 0.449 \\
  & \textbf{$R^2$}  & \textbf{0.965} & \textbf{0.953} & \textbf{0.932} & \textbf{0.938} \\
\midrule
\multirow{4}{*}{RVM}
  & $a$    & 0.875 & 0.911 & 0.945 & 0.960 \\
  & $b$    & 0.384 & 0.403 & 0.267 & 0.177 \\
  & $c$    & 1.062 & 0.827 & 0.810 & 0.874 \\
  & \textbf{$R^2$}  & \textbf{0.995} & \textbf{0.976} & \textbf{0.955} & \textbf{0.986} \\
\midrule
\multirow{4}{*}{LPVM}
  & $a$    & 0.876 & 0.912 & 0.944 & 0.961 \\
  & $b$    & 0.143 & 0.112 & 0.103 & 0.086 \\
  & $c$    & 1.066 & 0.984 & 0.844 & 0.812 \\
  & \textbf{$R^2$}  & \textbf{0.981} & \textbf{0.972} & \textbf{0.966} & \textbf{0.981} \\
\midrule
\multirow{4}{*}{LVVM}
  & $a$    & 0.874 & 0.911 & 0.944 & 0.960 \\
  & $b$    & 0.551 & 0.455 & 0.361 & 0.299 \\
  & $c$    & 1.175 & 1.051 & 0.943 & 0.881 \\
  & \textbf{$R^2$}  & \textbf{0.999} & \textbf{0.998} & \textbf{0.996} & \textbf{0.996} \\
\bottomrule
\end{tabular}
\caption{Fitted parameters of the stretched-exponential relationship for all models and system sizes $N$.}
\label{tab:strexp_params}
\end{table}

The original Vicsek model focuses on the critical behavior of the order parameter with respect to noise and density near the order-disorder transition, where power-law scalings are observed. In contrast, here we investigate how the final synchronization level depends on the initial average degree of the interaction network over a selected range of ${\langle d\rangle_i}$. Motivated by the saturation behaviour of $v_a$ at large ${\left\langle d \right\rangle _{i}}$, we expect the approach to the asymptotic value to follow an exponential-type law. Therefore we compare several candidate forms for the dependence of $1 - v_a({\left\langle d \right\rangle _{i}})$ on ${\left\langle d \right\rangle _{i}}$, including a simple exponential decay, a power-law decay and a stretched-exponential form. Power-law functions fail to reproduce the rapid initial decrease followed by a slow saturation, whereas exponential-type functions provide a much better description of the data. Among them, the stretched-exponential law offers the most accurate agreement across models and parameter sets,
\begin{equation}
v_a({\left\langle d \right\rangle _{i}}) = 1 -  a e^{-b ({\left\langle d \right\rangle _{i}})^c},
\label{eq:strexp_fit_N}
\end{equation}
where \(a\) quantifies the initial gap to full synchronization, \(b\) sets the typical degree scale on which synchronization builds up, and \(c\) controls how fast \(v_a\) approaches unity as ${\left\langle d \right\rangle _{i}}$ increases.

\begin{figure}[t]
    \centering
    \begin{minipage}[t]{0.57\textwidth}
        \vspace{0pt}
        \centering
        \includegraphics[width=\linewidth]{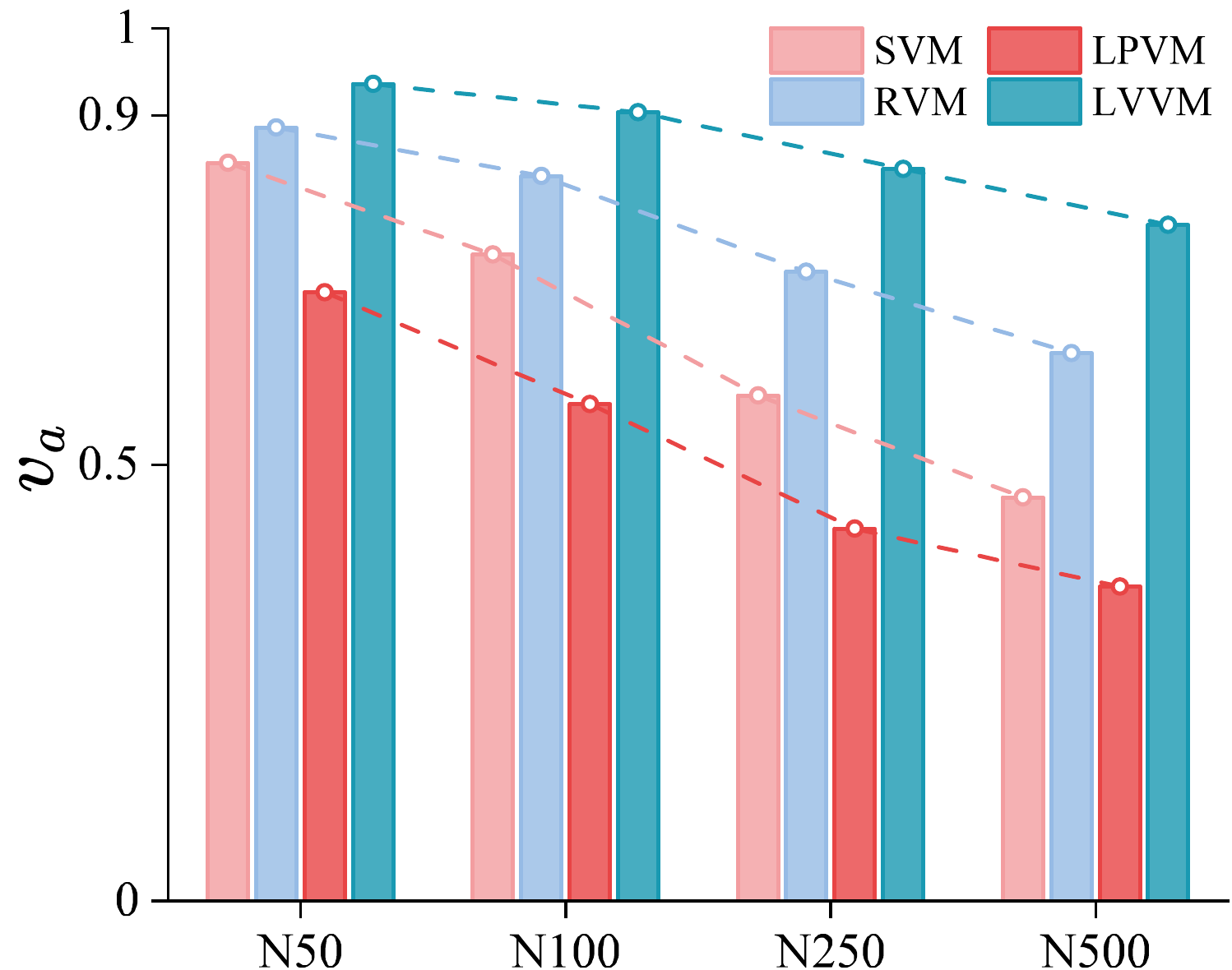}
    \end{minipage}%
    \hfill
    \begin{minipage}[t]{0.40\textwidth}
        \vspace{0pt}
        \centering
        \includegraphics[width=\linewidth]{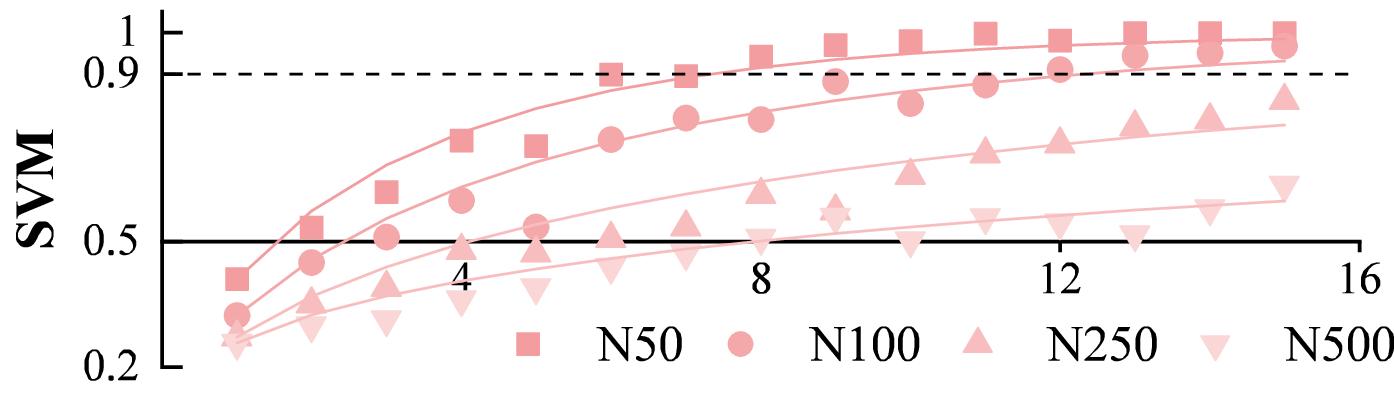}\\[2pt]
        \includegraphics[width=\linewidth]{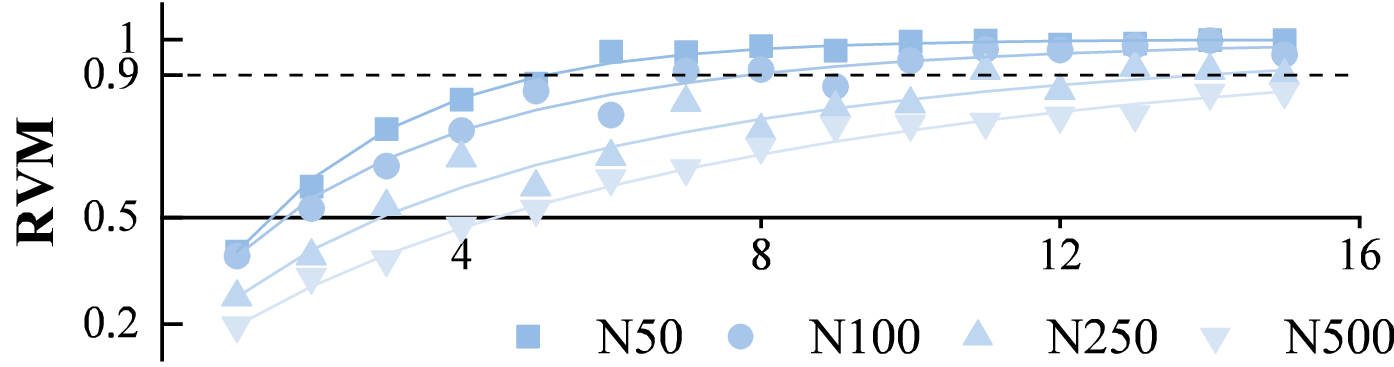}\\[2pt]
        \includegraphics[width=\linewidth]{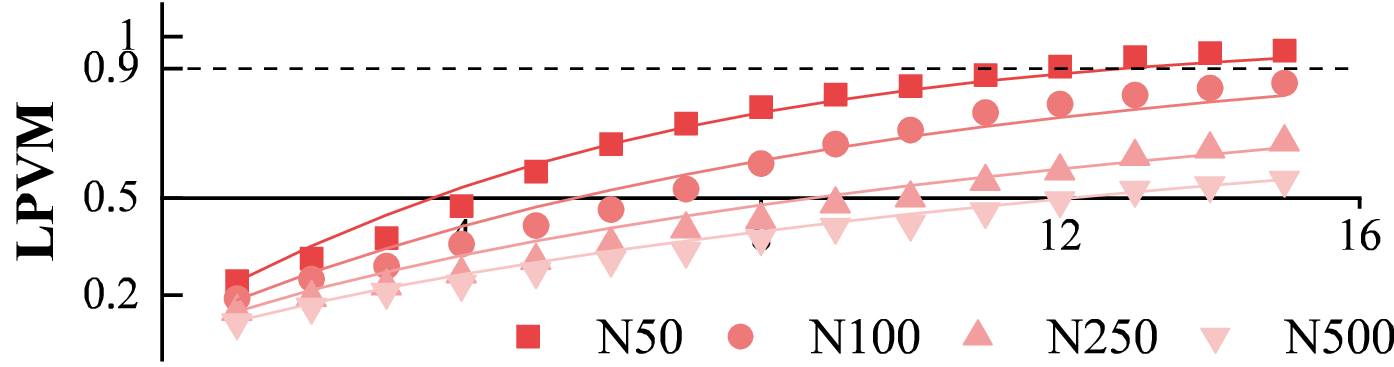}\\[2pt]
        \includegraphics[width=\linewidth]{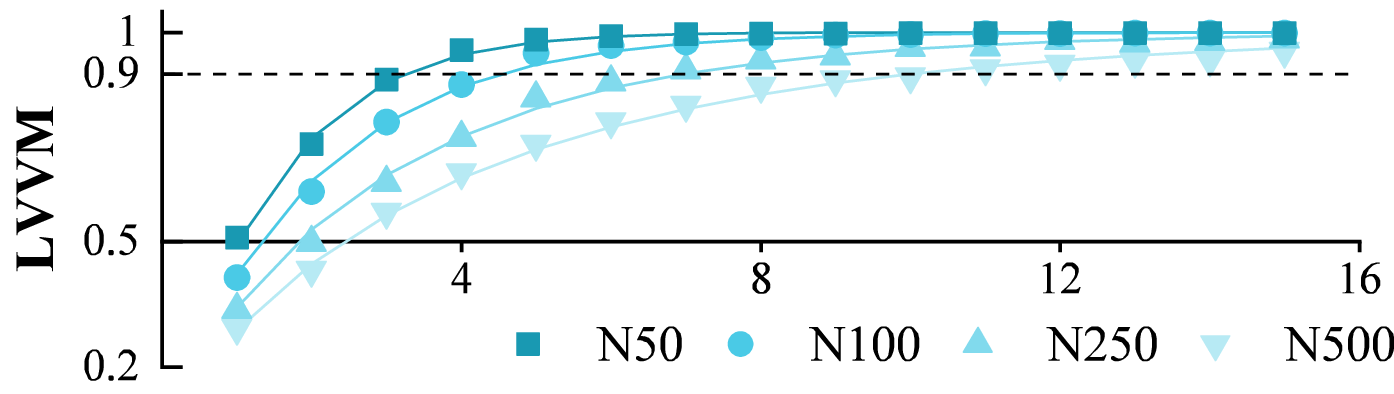}
    \end{minipage}
    \caption{Final synchronization level $v_a$ for the four models SVM, RVM, LPVM and LVVM. The left panel shows the dependence of $v_a$ on the system size $N$. The right panel shows the dependence of $v_a$ on the initial average degree $\langle d\rangle_{i}$ for each model, with symbols indicating $N = 50, 100, 250, 500$.}
    \label{fig:models_va_all}
\end{figure}

As shown in Fig.~\ref{fig:strexp_fits}, for each model and system size the stretched-exponential fits reproduce both the rapid initial decay and the subsequent slow saturation of $1 - v_a$ as the initial average degree $\langle d\rangle_i$ increases. LPVM and LVVM are reproduced almost perfectly over the whole range of $\langle d\rangle_i$, whereas SVM and RVM show slightly larger deviations from the fitted lines. In Tab.~\ref{tab:strexp_params}, all $R^2$ values are larger than $0.93$, and in most cases exceed $0.96$, confirming that the stretched-exponential form provides a quantitatively accurate description of the data. 

In summary, these numerical results clarify how the interaction topology and the initial connectivity jointly shape the collective dynamics of self-propelled particles. As shown in Fig.~\ref{fig:models_va_all}, the visibility-based LVVM displays the highest robustness to increases in system size and to agent heterogeneity. Moreover, the fact that all models are well described by a stretched-exponential dependence of the order parameter on the initial average degree indicates that they approach synchronization in a systematic way, despite their different microscopic rules and large-scale topologies. Combined with the ER-like and BA-like features, these findings provide a quantitative framework for predicting the asymptotic state of swarm systems from their initial connectivity and offer theoretical guidelines for the design and optimization of communication protocols in multi-agent networks.

\section{Conclusion}
\label{sec4}

In this study, we have proposed two novel Vicsek-type models (LPVM and LVVM) incorporating biological characteristics of heterogeneous sensing and influence, and have analyzed them alongside the established Standard and Restricted-view-angle models. By adopting a complex network perspective, we have revealed that while the homogeneous models generate ER like interaction graphs, our proposed heterogeneous variants naturally induce heavy-tailed, scale-free networks similar to the BA topology. A comparative analysis demonstrates that the visibility-based interaction rule in the LVVM confers significantly robustness for synchronization, highlighting that expanding the influence of leaders is more effective than enhancing their sensory range. Furthermore, we have identified a stretched-exponential scaling law that quantitatively links the initial network connectivity to the final synchronization level across all models. These findings provide a theoretical framework for predicting the asymptotic states of swarm systems and offer essential guidelines for the design of engineered multi-agent networks.

\section*{Acknowledgements}
This work was supported in part by the National Natural Science Foundation of China under Grants 62088101, and 62003015.

\section*{Code availability statement}
The code used in this study is available from the corresponding author (email: bit$\_$chen@bit.edu.cn) by request.

\end{document}